\documentclass[conference]{IEEEtran}
\IEEEoverridecommandlockouts
\usepackage{cite}
\usepackage{amsmath,amssymb,amsfonts}
\usepackage{algorithm}
\usepackage{algpseudocode}
\usepackage{graphicx}
\usepackage{subcaption}
\usepackage{bm}
\usepackage{pifont}
\usepackage[dvipsnames,svgnames]{xcolor, colortbl}
\usepackage{multirow}
\usepackage{threeparttable}
\usepackage{textcomp}
\usepackage{xcolor}
\usepackage{hyperref}
\hypersetup{%
	colorlinks=true,
	citecolor=black,
	linkcolor=black,
}
\newcommand{\ie}{\textit{i.e}}
\begin{document}
\bstctlcite{IEEEexample:BSTcontrol}

\title{DFT-s-OFDM for sub-THz Transmission - Tracking and Compensation of Phase Noise \\}

\author{\IEEEauthorblockN{Yaya Bello, Jean-Baptiste Doré, David Demmer}  \IEEEauthorblockA{CEA-Leti, Univ. Grenoble Alpes, F-38000 Grenoble, France}

\{yaya.bello, jean-baptiste.dore, david.demmer\}@cea.fr}

\maketitle

\begin{abstract}
For future wireless communication technologies, an increase in capabilities such as throughput is strongly expected. Transmission in the sub-THz bands ($>$90 GHz) seems to be the potential solution to meet the ever-increasing capacity demands due to the large unexploited bandwidth. Oscillators used at these frequencies generate phase noise that induces critical distortions in the signal that must be addressed. The correlated nature of PN makes it difficult to overcome. Nowadays, there is a growing interest in considering the extension of multicarrier based waveforms of the 5G new radio for transmissions in the sub-THz bands. In this paper, we introduce a new algorithm called the interpolation filter (IF), which efficiently estimates and compensates PN effects on DFT-s-OFDM systems. Specifically, it is based on the use of stochastic properties of the PN and is compatible with the 3GPP phase tracking reference signal scheme. We highlight a performance improvement over known techniques when using high-order modulation.
\end{abstract}

%%%%%%%%%%%%%%%%%%%%%%%%%%%%%%%%%%%%%%%%%%%%%%%%%%%%%%%%%%%%%%%%%%%%%%%%%%%%%%%%%%
\section{Introduction}
%%%%%%%%%%%%%%%%%%%%%%%%%%%%%%%%%%%%%%%%%%%%%%%%%%%%%%%%%%%%%%%%%%%%%%%%%%%%%%%%%%
To achieve a throughput of $1$ Terabits per second with a wireless system, new frequency bands with extremely wide bandwidth must be used. One possible solution discussed in the literature is to transmit in the so-called sub-THz bands ($>$90 GHz)\cite{Petrov} where it is possible to increase the bandwidth of the signal \cite{Dore2018}. One of the main concerns in using these bands is the ability to design a radio frequency front end with good phase noise (PN) properties. The PN generated by oscillators at these frequencies has strong impacts on communication systems. It is therefore crucial to address phase noise impairments in order to be able to standardize the sub-THz bands for wireless communications. 

The 3rd Generation Partnership Project (3GPP) focuses on the implementation of 5G-New Radio (5G-NR) multicarrier waveforms in frequencies above $90$GHz \cite{Tervo}. DFT-s-OFDM represents a potential candidate for technologies beyond 5G. Indeed, it has a lower peak-to-average power ratio (PAPR) \cite{Berardinelli} unlike OFDM. Thus, it allows to operate power amplifiers more efficiently. However, the PN deteriorates its performance as in OFDM (rotation of received signals and inter-carrier interference (ICI) \cite{Petrovic}).

Phase tracking reference signal insertion (PTRS) is the method used in 5G-NR to track the phase of the local oscillator generated at the transmitter and receiver \cite{Qi}. The authors of \cite{Afshang} proposed two methods to remove the ICI effects caused by PN for OFDM systems. Low and high complexity PN estimation algorithms are discussed in \cite{Sibel} for DFT-s-OFDM working at mm-wave frequencies. In \cite{Bhatti}, the authors proposed the discrete cosine transform (DCT) algorithm for PN estimation. They showed that good performance can be obtained for a single carrier waveform depending on the number of DCT coefficients used to estimate the PN.  

Although many contributions have proposed estimation algorithms to compensate the PN, there is no study on PN estimation algorithms that consider the use of its stochastic properties in DFT-s-OFDM system. In this paper, we propose an estimation algorithm based on the linear minimum mean square error (LMMSE). We compare it to existing algorithms. 
\smallskip

Regarding the above considerations, the contributions of this paper are as follows:
\begin{itemize}
    \item We present an algorithm for the estimation of PN induced effects: \textbf{interpolation filter} (IF). It considers stochastic properties of all interference terms induced by the PN on the system. It represents the main contribution of this study.
    
    \item We show that the IF algorithm is suitable to contiguous PTRS and is compatible with 3GPP PTRS pattern distribution. We highlight that it outperforms other presented algorithms in the case of low PTRS density which leads to better spectrum efficiency.
    
    \item Performance analysis are provided in terms of bit error rate (BER) and transport block error rate (TBLER) for high order modulation schemes and using the 5G-NR low density parity check (LDPC) code.
\end{itemize}

\subsection*{Notations}
In what follows, underlined bold lowercase letters $\underline{\mathbf{a}}$ indicate column vectors, with $a_k$ denoting the $k^{th}$ element of the column vector. Higher boldface letters $\mathbf{A}$ denote a matrix. The symbols $\vert \cdot \vert $, $\text{arg}(\cdot)$, $(\cdot)^T$, $(\cdot)^*$ and $(\cdot)^H$ respectively denote the magnitude value, the phase value, the transpose, the conjugate and the transpose-conjugate. The matrix $\mathbf{A}^\dagger$ which is defined as $\mathbf{A}^\dagger=\left(\mathbf{A^H}\mathbf{A}\right)^{-1}\mathbf{A^H}$ denotes the Moore-Penrose pseudo-inverse of matrix $\mathbf{A}$. Underlined boldface numbers $\underline{\mathbf{N}}_X$ indicate column vectors of size $X$ which contain the number $\mathbf{N}$ in all rows. The operator $\odot$ represents the Hadamard product. 

%\begin{figure}[tb]
    %\centering
    %\includegraphics[width=1\columnwidth]{./figs/SM3.JPG}
    %\caption{DFT-s-OFDM transmission chain. \vspace{-2cm}}
    %\label{fig:SM}
%\end{figure}

%%%%%%%%%%%%%%%%%%%%%%%%%%%%%%%%%%%%%%%%%%%%%%%%%%%%%%%%%%%%%%%%%%%%%%%%%%%%%%%%%%
\section{System Model}\label{SysMod}
%%%%%%%%%%%%%%%%%%%%%%%%%%%%%%%%%%%%%%%%%%%%%%%%%%%%%%%%%%%%%%%%%%%%%%%%%%%%%%%%%%
In this section, we derive analytical expression of the received signal considering a DFT-s-OFDM system in the presence of PN.

%%%%%%%%%%%%%%%%%%%%%%%%%%%%%%%%%%%%%%%%%%%%%%%%%%%%%%%%%%%%%%%%%%%%%%%%%%%%%%%%%%
\subsection{Channel Model}\label{PN}
%%%%%%%%%%%%%%%%%%%%%%%%%%%%%%%%%%%%%%%%%%%%%%%%%%%%%%%%%%%%%%%%%%%%%%%%%%%%%%%%%%
It has been demonstrated that in sub-THz frequencies, the line-of-sight (LoS) provides most of the energy contribution \cite{Pometcu}. It is obtained by using high gain directional antennas which spatially filter the channel. For this work, we assume a LoS propagation channel considered as frequency flat. We also assume a coherent communication system impaired by PN with a receiver synchronized in time and frequency. We consider the PN model used by 3GPP \cite{3GPP1}. For this model, the power spectral density (PSD) expressions of PN are provided for both transmitter and receiver.

%%%%%%%%%%%%%%%%%%%%%%%%%%%%%%%%%%%%%%%%%%%%%%%%%%%%%%%%%%%%%%%%%%%%%%%%%%%%%%%%%
\subsection{DFT-s-OFDM in the Presence of Phase Noise}
%%%%%%%%%%%%%%%%%%%%%%%%%%%%%%%%%%%%%%%%%%%%%%%%%%%%%%%%%%%%%%%%%%%%%%%%%%%%%%
We consider a discrete-time DFT-s-OFDM system with $N_p$ subcarriers for the OFDM modulation. We assume $N_a$ active subcarriers with $N_a<N_p$ for the discret Fourier transform (DFT) spreading. We denote $s_k$ as the transmitted symbol at the DFT subcarrier $k$ with $k = \{0, \cdots ,N_a-1\}$. Symbols $s_k \in \mathcal{C}$ where $\mathcal{C}$ is the set of the selected modulation scheme.

Considering the channel model presented above, the discrete-time received signal after the fast Fourier transform (FFT) and inverse DFT (IDFT) at the subcarrier index $k$ for each DFT-s-OFDM symbol is:
\begin{equation}
 \begin{aligned}
    r_k = &  s_k.\alpha_k + \beta_k + w_k,
 \label{eq_stilde}
 \end{aligned}
\end{equation}
where $w_k \backsim \mathbb{C}\mathcal{N}(0, \sigma_n^2)$ represents thermal noise after the FFT and IDFT operations. $\alpha_k$ and $\beta_k$ are respectively the terms responsible of the rotation of transmitted symbols and the ICI. They are defined as follows:
\begin{equation}
     \alpha_k =\frac{1}{N_a}\frac{1}{N_p}A_k, \quad \quad \beta_k = \frac{1}{N_a}\frac{1}{N_p}\sum\limits_{n=0,n\ne k}^{N_a-1}s_n B_{n,k},
\end{equation}  
\begin{equation}
    A_k = \sum\limits_{f=0}^{N_a-1}\sum\limits_{m=0}^{N_p-1} \sum\limits_{p=0}^{N_p-1}e^{j2\pi\frac{(m-f)}{N_p}p} \notag e^{-j2\pi\frac{(m-f)}{N_a}k}e^{j\phi(p)},
    \label{eq_A}
\end{equation}
\begin{equation}
    B_{n,k} = \sum\limits_{f=0}^{N_a-1}\sum\limits_{m=0}^{N_p-1}\sum\limits_{p=0}^{N_p-1}e^{j2\pi\frac{(m-f)}{N_p}p}e^{j2\pi\frac{kf-nm}{N_a}}e^{j\phi(p)}\cdot \notag
 \end{equation}
We have different values of $\alpha_k, \beta_k$ and $w_k$ on each DFT-s-OFDM received symbols. The term $\phi$ in $A_k$ and $B_{n,k}$ represents the discrete-time PN after sampling at the reception. It results from the sum of the PN at the transmitter and receiver. We suppose that the term $\alpha_k$ can be expressed as $\alpha_k \simeq e^{j\phi'_k}$, therefore the system model expression in (\ref{eq_stilde}) becomes:
 \begin{align}
    r_k \simeq & s_k.e^{j\phi'_k} + \beta_k + w_k.
    \label{eq_R}
 \end{align}

%%%%%%%%%%%%%%%%%%%%%%%%%%%%%%%%%%%%%%%%%%%%%%%%%%%%%%%%%%%%%%%%%%%%%%%%%%%%%%%%%%
\subsection{PTRS Pattern}\label{PTRS_Pat}
%%%%%%%%%%%%%%%%%%%%%%%%%%%%%%%%%%%%%%%%%%%%%%%%%%%%%%%%%%%%%%%%%%%%%%%%%%%%%%%%%%
PTRS are assumed to be distributed among the time/frequency grid as shown in \mbox{Fig. \ref{fig:PTRS}}. PTRS are inserted in each DFT-s-OFDM symbol and before the DFT spreading operation to avoid the increase of the PAPR. $K$ and $N_{DFT}$ are respectively the number of PTRS in a DFT-s-OFDM symbol and the number of transmitted DFT-s-OFDM symbols. $\chi_p = \{p_0,p_1,\cdots p_{K-1} \}$ denotes the set that contains all the DFT-s-OFDM subcarrier indexes where PTRS symbols are positioned. Considering the PTRS distribution in Fig. \ref{fig:PTRS}-(a), $p_0 = 0$ so for the rest of the paper,  $\chi_p = \{0,p_1,\cdots p_{K-1}\}$.

\vspace{-0cm}
%%%%%%%%%%%%%%%%%%%%%%%%%%%%%%%%%%%%%%%%%%%%%%%%%%%%%%%%%%%%%%%%%%%%%%%%%%%%%%%%%%
\section{Phase Noise Estimation Algorithms}\label{Algo}
%%%%%%%%%%%%%%%%%%%%%%%%%%%%%%%%%%%%%%%%%%%%%%%%%%%%%%%%%%%%%%%%%%%%%%%%%%%%%%%%%%
In this section, we present and describe five estimation techniques: common phase error (CPE) estimation \cite{Sibel}, linear interpolation \cite{Sibel} and DCT \cite{Bhatti} algorithms are parts of the literature. A simple and low complexity algorithm called "constant interpolation" is also presented. The last algorithm described represents the contribution of this paper.

%%%%%%%%%%%%%%%%%%%%%%%%%%%%%%%%%%%%%%%%%%%%%%%%%%%%%%%%%%%%%%%%%%%%%%%%%%%%%%%%%%
\subsection{State-of-the-art Estimation Algorithms}
%%%%%%%%%%%%%%%%%%%%%%%%%%%%%%%%%%%%%%%%%%%%%%%%%%%%%%%
%%%%%%%%%%%%%%%%%%%%%%%%%%%%%
\subsubsection{CPE Estimation (CPEE)}
%%%%%%%%%%%%%%%%%%%%%%%%%%%%%%%%%%%%%%%%%%%%%%%%%%%%%%%%%%%%%%%%%%%%%%%%%%%%%%%%%%
It consists in estimating the average phase error over each DFT-s-OFDM symbol. Let $\hat{\phi}_{CPEE}$ be the average value obtained as follows: 
\begin{equation}
     \hat{\phi}_{CPEE} = \text{arg}\left (\sum\limits_{i \in \chi_p }  \frac{r_is^*_i}{\vert s_i \vert^2}\right)\cdot
     \label{Eq_CPEE}
\end{equation}
According to  the transmitted symbol $s_n$, the estimated symbol $\hat{s}_n$ for any DFT-s-OFDM block is:
\begin{equation}
     \hat{s}_n = r_n.e^{-j\hat{\phi}_{CPEE}}, \quad \forall n \in [0,N_a-1]\cdot
     \label{Eq_Est_CPEE}
\end{equation}
%\smallskip
\begin{figure}[tb]
    \centering
    \includegraphics[width=1.1\columnwidth]{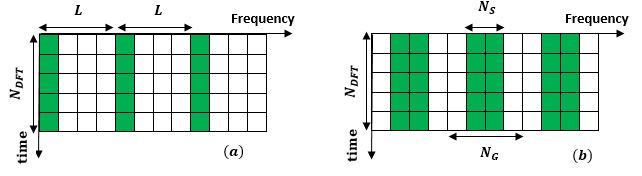}
    \caption{(a): Distributed PTRS pattern $N_{DFT}= 5$, $L = 4$, $N_a = 12$ and $K = 3$. (b): Contiguous PTRS pattern $N_{DFT}= 5$, $N_a = 12$, $N_G = 3$ and $N_S = 2$. \vspace{-0.5cm}}
    \label{fig:PTRS}
\end{figure}
%%%%%%%%%%%%%%%%%%%%%%%%%%%%%%%%%%%%%%%%%%%%%%%%%%%%%%%%%%%%%%%%%%%%%%%%%%%%%%%%%%
\subsubsection{Linear Interpolation (LI)}
%%%%%%%%%%%%%%%%%%%%%%%%%%%%%%%%%%%%%%%%%%%%%%%%%%%%%%%%%%%%%%%%%%%%%%%%%%%%%%%%%%
It is based on making a linear interpolation from received PTRS symbols for each DFT-s-OFDM symbol \cite{Sibel}. The estimated PN is obtained as follows:
\begin{equation}
     \underline{\bm{\hat{\phi}}}_{\mathbf{\textit{LI}}} = \underset{N_a}{\mathcal{LI}}\left (\underline{\bm{\hat{\phi}}}_{\chi_p} \right),
     \label{Eq_LI2}
\end{equation}
where $\underset{N_a}{\mathcal{LI}}$ denotes the linear interpolation operation of size $N_a$. $\underline{\bm{\hat{\phi}}}_{\chi_p}$ is the vector which contains all the $\{\hat{\phi}_i\}_{i \in \chi_p}$ values. The estimated vector $\underline{\mathbf{\hat{s}}}$ is expressed as follows:
\begin{equation}
     \underline{\mathbf{\hat{s}}} = \underline{\mathbf{r}}\odot e^{-j\underline{\bm{\hat{\phi}}}_{\mathbf{\textit{LI}}}},
     \label{Eq_Est_LI}
\end{equation}
where $\underline{\mathbf{\hat{s}}}$ and $\underline{\mathbf{r}}$ are respectively the vector that contains all the estimated $\hat{s}_n$ and received $r_n$ symbols with $n = [0,N_a-1]$.
\smallskip

%%%%%%%%%%%%%%%%%%%%%%%%%%%%%%%%%%%%%%%%%%%%%%%%%%%%%%%%%%%%%%%%%%%%%%%%%%%%%%%%%%
\subsubsection{Discrete Cosine Transform (DCT)}
%%%%%%%%%%%%%%%%%%%%%%%%%%%%%%%%%%%%%%%%%%%%%%%%%%%%%%%%%%%%%%%%%%%%%%%%%%%%%%%%%%
This estimation algorithm relies on the approximation of PN by discrete cosine transform (DCT) basis functions. This is possible by determining the number of DCT coefficients that will be sufficient to correctly estimate the PN. The estimated PN is expressed as follows:
\begin{equation}
     \underline{\bm{\hat{\phi}}} = \phi_{av}\underline{\mathbf{1}}_{N_a} + \bm{\Psi}_{N_a}\underline{\hat{\mathbf{x}}}_\mathbf{p},
     \label{eq_DCT}
\end{equation}
\begin{equation}
\phi_{av} = \text{arg}( \underset{i \in \chi_p}{\sum}r_i) \text{ and~} \underline{\hat{\mathbf{x}}}_\mathbf{p} = (\bm{\Psi}^T_{K}\bm{\Psi}_{K})^{-1}\bm{\Psi}^T_{K}\underline{\hat{\mathbf{r}}}'_\mathbf{p},
\label{eq_DCT2}
\end{equation}
where $\underline{\hat{\mathbf{r}}}'_\mathbf{p}$ represents the vector which contains all the $r'_i$ such that $\forall i \in \chi_p$ :
\begin{equation}
     r'_i = \text{arg}\left( r_i s^*_i e^{-j\phi_{av}}\right)\cdot
     \label{eq_DCT3}
\end{equation}
The matrix $\bm{\Psi}_{N_a}$ contains all the $\psi_n(k)$ where $\psi_n(k)$ represents the orthonormal DCT basis function \cite{Bhatti}. The matrix $\bm{\Psi}_{K}$ contains all $\psi_n(p)$ with $p = \{0, \cdots,K-1\}$. For both $\psi_n(k)$ and $\psi_n(p)$, $n = \{0,\cdots,N_D-1\}$ with $N_D$ the number of DCT coefficients. For the inversion operation in (\ref{eq_DCT2}) to be possible, it is necessary that:
\begin{equation}
     N_D \leq K\cdot
     \label{eq_DCT4}
\end{equation}

%%%%%%%%%%%%%%%%%%%%%%%%%%%%%%%%%%%%%%%%%%%%%%%%%%
\subsection{Constant Interpolation (CI)}
%%%%%%%%%%%%%%%%%%%%%%%%%%%%%%%%%%%%%%%%%%%%%%%%%%%
It considers that the PN is constant between consecutive PTRS. Let $\hat{\phi}_i$ be the estimated PN at the $i^{th}$ PTRS index in a DFT-s-OFDM symbol such that $i \in \chi_p$ and defined as:
\begin{equation}
     \hat{\phi}_i = \underset{i \in \chi_p}{\text{arg}}\left (r_is^*_i\right)\cdot
     \label{Eq_LI}
\end{equation}
The estimated PN $\hat{\phi}_n$ is expressed as follows:
\begin{align}
  \hat{\phi}_n = \left\{
    \begin{array}{ll}
       \hat{\phi}_{0} \quad  \quad \quad \text{when} \; \; \; \quad  0 \leq n \leq p_1-1\\
       \hat{\phi}_{p_1} \quad  \quad \; \; \text{when} \quad \; \; p_1 \leq n \leq p_2-1\\
       \quad \vdots \quad  \quad \quad \quad  \quad \quad \quad \quad \quad \; \vdots \\
       \hat{\phi}_{p_{K-1}} \quad \; \text{when} \; \; p_{K-1} \leq n \leq N_a-1 \cdot\\
    \end{array}
\right.
\end{align}
The estimated symbol $\hat{s}_n$ is obtained by computing the expression (\ref{Eq_Est_CPEE}).

%%%%%%%%%%%%%%%%%%%%%%%%%%%%%%%%%%%%%%%%%%%%%%%%%%%
%%%%%%%%%%%%%%%%%%%%%%%%%%%%%%%%%%%%%%%%%%%%%%%%%%%%%%%%%%%%%%%%%%%%%%%%%%%%%%%%%%
\subsection{Proposed Estimation Algorithm : \textbf{Interpolation Filter} (IF)}
%%%%%%%%%%%%%%%%%%%%%%%%%%%%%%%%%%%%%%%%%%%%%%%%%%%%%%%
%%%%%%%%%%%%%%%%%%%%%%%%%%%%%%%%%%%%%%%%%%%%%%%%%%%%%%%%%%%%%%%%
IF algorithm relies on the use of stochastic properties of the PN based on its correlated nature. According to the received signal expression in (\ref{eq_R}), the new PN seen at the receiver is $\phi'_k$. Therefore, the objective is to find a matrix $\mathbf{Z}$ such that:
\begin{align}
    &\underset{\mathbf{Z}}{\text{min}}\; E\left[ \lVert \mathbf{Z} \underline{\mathbf{a}}_\mathbf{p} - \underline{\mathbf{\Phi}}'\rVert^2 \right] \label{eq_Z}\\ &= E\left [\mathcal{T}_r \lbrace \mathbf{Z}\underline{\mathbf{a}}_\mathbf{p}\underline{\mathbf{a}}_\mathbf{p}^H\mathbf{Z}^H - \mathbf{Z}\underline{\mathbf{a}}_\mathbf{p}\underline{\mathbf{\Phi}}'^H - \underline{\mathbf{\Phi}}'\;\underline{\mathbf{a}}_\mathbf{p}^H\mathbf{Z}^H + \underline{\mathbf{\Phi}}'\;\underline{\mathbf{\Phi}}'^H \rbrace \right],\notag
\end{align}
where $\mathbf{Z}$ is the interpolation filter of size $N_a \times K$. The vector $\underline{\mathbf{a}}_\mathbf{p}$ contains all the values obtained from the inserted PTRS and $\underline{\mathbf{\Phi}}'=[e^{j\phi'_0},\cdots,e^{j\phi'_{N_a-1}}]^T$ is the PN vector that we want to estimate. Finding the interpolation matrix $\mathbf{Z}$ that fills (\ref{eq_Z}) is equivalent to solving:
\begin{align}
    &\frac{\partial}{\partial \mathbf{Z}} E\left[ \lVert \mathbf{Z} \underline{\mathbf{a}}_\mathbf{p} - \underline{\mathbf{\Phi}}'\rVert^2 \right] = 0 \notag \\
    & \Longleftrightarrow E\left[ \underline{\mathbf{a}}_\mathbf{p}\underline{\mathbf{a}}_\mathbf{p}^H\mathbf{Z}^H \right] = E\left[\underline{\mathbf{a}}_\mathbf{p}\underline{\mathbf{\Phi}}'^H \right]\cdot  
    \label{eq_Drv2}
\end{align}
where $\frac{\partial}{\partial \mathbf{Z}}$ is the derivative operation with respect to $\mathbf{Z}$ and $\mathcal{T}_r \lbrace \cdot \rbrace$ denotes the trace operation of the argument. The next step is to express $\underline{\mathbf{a}}_\mathbf{p}$ depending on our system model.
From (\ref{eq_R}), we define $a_i$ such that:
\begin{equation}
    a_i = \frac{r_is^*_i}{\vert s_i \vert^2} = e^{j\phi'_i} + \frac{\beta_is^*_i}{\vert s_i \vert^2} + \frac{w_is^*_i}{\vert s_i \vert^2}, \quad \forall i \in \chi_p 
    \label{eq_AK}
\end{equation}
where $r_i$ and $s_i$ are respectively the received and the inserted PTRS symbol. Let $\underline{\mathbf{a}}_\mathbf{p}=[a_{0},a_{p_1},\cdots, a_{p_{K-1}}]^T$ be the vector which contains all $\{a_i\}_{i \in \chi_p}$. According to (\ref{eq_AK}), we obtain: 
\begin{equation}
 \underline{\mathbf{a}}_\mathbf{p} = \mathbf{M_p}\underline{\mathbf{\Phi}}' + \mathbf{M_p}\underline{\bm{\beta}} \odot
\underline{\mathbf{s}}_\mathbf{p} + \mathbf{M_p}\underline{\mathbf{w}} \odot
\underline{\mathbf{s}}_\mathbf{p},
\label{eq_AP33}
\end{equation}
where $\mathbf{M_p}$ denotes the sampling matrix of the PTRS values. The vector $\underline{\bm{\beta}} = [\beta_0,\cdots,\beta_{N_a-1}]^T$, $\underline{\mathbf{w}}=[w_0,\cdots,w_{N_a-1}]^T$ and $\underline{\mathbf{s}}_\mathbf{p}$ is defined as follows:
\begin{equation}
    \underline{\mathbf{s}}_\mathbf{p} = \left[\frac{s^*_{0}}{\vert s_{0} \vert^2},\cdots, \frac{s^*_{p_{K-1}}}{\vert s_{p_{K-1}} \vert^2}\right]^T\cdot \notag
\end{equation}
Let's denote ${\mathcal{R}_{\mathbf{\Phi'}} = E\left[\underline{\mathbf{\Phi}}' \underline{\mathbf{\Phi}}'^H\right]}$, ${\mathcal{R}_{\bm{\beta}} = E\left[ \underline{\bm{\beta}}\;\underline{\bm{\beta}} ^H\right]}$ and $\mathcal{R}_{\mathbf{w}} = E\left[ \underline{\mathbf{w}}\;\underline{\mathbf{w}}^H\right]$ as the covariance matrix respectively of $\underline{\mathbf{\Phi}}'$, $\underline{\bm{\beta}}$ and $\underline{\mathbf{w}}$. Considering that ${E\left[ \underline{\mathbf{w}}\right]=\mathbf{0}_{N_a}}$ because ${w_k \backsim \mathbb{C}\mathcal{N}(0, \sigma_n^2)}$ and taking the hypothesis that ${E\left[\; \underline{\bm{\beta}} \;\right]=\mathbf{0}_{N_a}}$, the interpolation matrix $\mathbf{Z}$ using (\ref{eq_Drv2})-(\ref{eq_AP33}) is expressed as follows:
\begin{equation}
 \mathbf{Z} = \mathcal{R}_{\mathbf{\Phi'}}^H\mathbf{M_p}^H\mathcal{A}^{\dagger},
\label{eq_Z2}
\end{equation}
with $\mathcal{A} = \mathcal{P} + \mathcal{Q} + \mathcal{V}$ where:
\begin{equation}
\begin{split}
 &\mathcal{P} = \mathbf{M_p}\mathcal{R}_{\mathbf{\Phi'}}^H\mathbf{M_p}^H, \mathcal{Q} = \mathbf{M_p}\mathcal{R}_{\bm{\beta}}^H\mathbf{M_p}^H\odot \underline{\mathbf{s}}_\mathbf{p}^H\underline{\mathbf{s}}_\mathbf{p} \\
 &\text{and } \mathcal{V} = \mathbf{M_p}\mathcal{R}_{\mathbf{w}}^H\mathbf{M_p}^H\odot \underline{\mathbf{s}}_\mathbf{p}^H\underline{\mathbf{s}}_\mathbf{p}\cdot 
 \end{split}
\label{eq_Z3}
 \end{equation}
The expression of the estimated PN $\underline{\bm{\hat{\phi}}}_{\mathbf{\textit{IF}}}$ vector is:
\begin{equation}
     \underline{\bm{\hat{\phi}}}_{\mathbf{\textit{IF}}} = \text{arg}\left( \mathbf{Z}\underline{\mathbf{a}}_{\mathbf{p}}  \right),
 \label{eq_ZPN}
 \end{equation}
and then the phase can be corrected by computing the expression (\ref{Eq_Est_LI}). 

%%%%%%%%%%%%%%%%%%%%%%%%%%%%%%%%%%%%%%%%%%%%%%%%%%%%%%%%%%%%%%%%%%%
\subsection{Discussions}
%%%%%%%%%%%%%%%%%%%%%%%%%%%%%%%%%%%%%%%%%%%%%%%%%%%%%%%%%%%%%%%%%%%%
The IF algorithm can also be implemented in both single-carrier and OFDM systems. For OFDM, the implementation of the IF algorithm corresponds to estimating and compensating the CPE. Different options can be considered for the estimation of the interpolation filter $\mathbf{Z}$. A first method consists in estimating online the covariance matrices based on the inserted PTRS. Once the matrices are estimated, the filter computation can be performed. A second option is to precompute the filter for a given PN range. This approach will be less efficient but less demanding in terms of computation. Finally, a last approach can be to compute a set of PN features. Then, we search among the known characteristics, the one that is the closest to our system and we use the corresponding filter. It should be noted that the matrix $\mathbf{Z}$ is usually computed for a given PN model. Thus, the filtering operation can be reduced to only a few multiplications for the phase estimation. 

%%%%%%%%%%%%%%%%%%%%%%%%%%%%%%%%%%%%%%%%%%%%%%%%%%%%%%%%%%%%
\begin{table}[tb]
\caption{Simulation parameters}
\centering
\begin{tabular}{llr}
\rowcolor[HTML]{EFEFEF} 
Sampling Frequency        & \multicolumn{1}{c}{\cellcolor[HTML]{EFEFEF}$F_s$} & $1966.08$ MHz                                                         \\
Numerology                & \multicolumn{1}{c}{$\mu$}                         & $6$                                                          \\
\rowcolor[HTML]{EFEFEF} 
Signal bandwidth          & \multicolumn{1}{c}{\cellcolor[HTML]{EFEFEF}$B$}   & $983.04$ MHz                                                         \\
Carrier frequency         & $F_c$                                             & \multicolumn{1}{r}{$140$ and $300$ GHz}                         \\
\rowcolor[HTML]{EFEFEF} 
Phase noise model         & \multicolumn{2}{r}{\cellcolor[HTML]{EFEFEF}3GPP \cite{3GPP1}}                                                \\
Tx IFFT size        & $N$                                               & \multicolumn{1}{r}{$2048$} \\
\rowcolor[HTML]{EFEFEF} 
Number of active carriers & $N_a$                                             & \multicolumn{1}{r}{\cellcolor[HTML]{EFEFEF}$1024$}                    \\
LDPC coding rate  & $R$    & \multicolumn{1}{r}{$0.7$} \\

\rowcolor[HTML]{EFEFEF} LDPC decoder iteration  &  $n_{it}$   & \multicolumn{1}{r}{\cellcolor[HTML]{EFEFEF} $15$} \\
LDPC decoder  &     & \multicolumn{1}{r}{Layered Min-Sum} \\

\rowcolor[HTML]{EFEFEF} Modulation  &     & \multicolumn{1}{r}{\cellcolor[HTML]{EFEFEF} $16, 64$ and $256$ QAM} \\
\\
\end{tabular}
\label{tab:simu}
\vspace{-0.8cm}
\end{table}

\vspace{-0.05cm}
%%%%%%%%%%%%%%%%%%%%%%%%%%%%%%%%%%%%%%%%%%%%%%%%%%%%%%%%%%%%
%%%%%%%%%%%%%%%%%%%%%%%%%%%%%%%%%%%%%%%%%%%%%%%%%%%%%%%%%%%%%%%%%%%%%%%%%%%%%%%%%%
\section{Numerical Results}\label{Res}
%%%%%%%%%%%%%%%%%%%%%%%%%%%%%%%%%%%%%%%%%%%%%%%%%%%%%%%%%%%%%%%%%%%%%%%%%%%%%%%%%%
In this section, we assess the performance of the estimation algorithms described previously. The parameters used for numerical results are depicted in TABLE \ref{tab:simu}. The discrete-time PN is generated by considering the method presented in \cite{3GPP2}. We assume a sampling period $T_s = 1/F_s$. For the proposed algorithm, it is necessary for the receiver to estimate the covariance matrices $\mathcal{R}_{\mathbf{\Phi'}}$, $\mathcal{R}_{\bm{\beta}}$ and $\mathcal{R}_{\mathbf{w}}$. This is achieved by means of a training sequence. DCT-$N_D$ means the implementation of the DCT algorithm with $N_D$ DCT coefficients and for numerical results, we arbitrarily choose \mbox{$N_D$ = $\{1,2,10,30\}$}.

%%%%%%%%%M%%%%%%%%%%%%%%%%%%%%%%%%%%%%%%%%%%%%%%%%%%%%%%%%%%%%%%%%%%%%%%%%%%%%%%%%
\subsection{Uncoded BER Performance: \textbf{Contiguous PTRS}}
%%%%%%%%%%%%%%%%%%%%%%%%%%%%%%%%%%%%%%%%%%%%%%%%%%%%%%%%%%%%%%%%%%%%%%%%%%%%%%%%%%
Here, we assume a 3GPP PTRS scheme as presented in Fig.~\ref{fig:PTRS}-(b): we divide each DFT-s-OFDM symbol in $N_G$ groups and every group contains $N_S$ PTRS. For contiguous PTRS, the number of inserted PTRS is ${K = N_G N_S}$. For both IF and DCT algorithms, we consider all PTRS symbols of each group. For CI, CPEE and LI algorithms, we consider only one estimated value per group which is obtained by averaging all the PTRS in each group.

\begin{figure}[tb]
\centering
    \includegraphics[width=0.98\columnwidth]{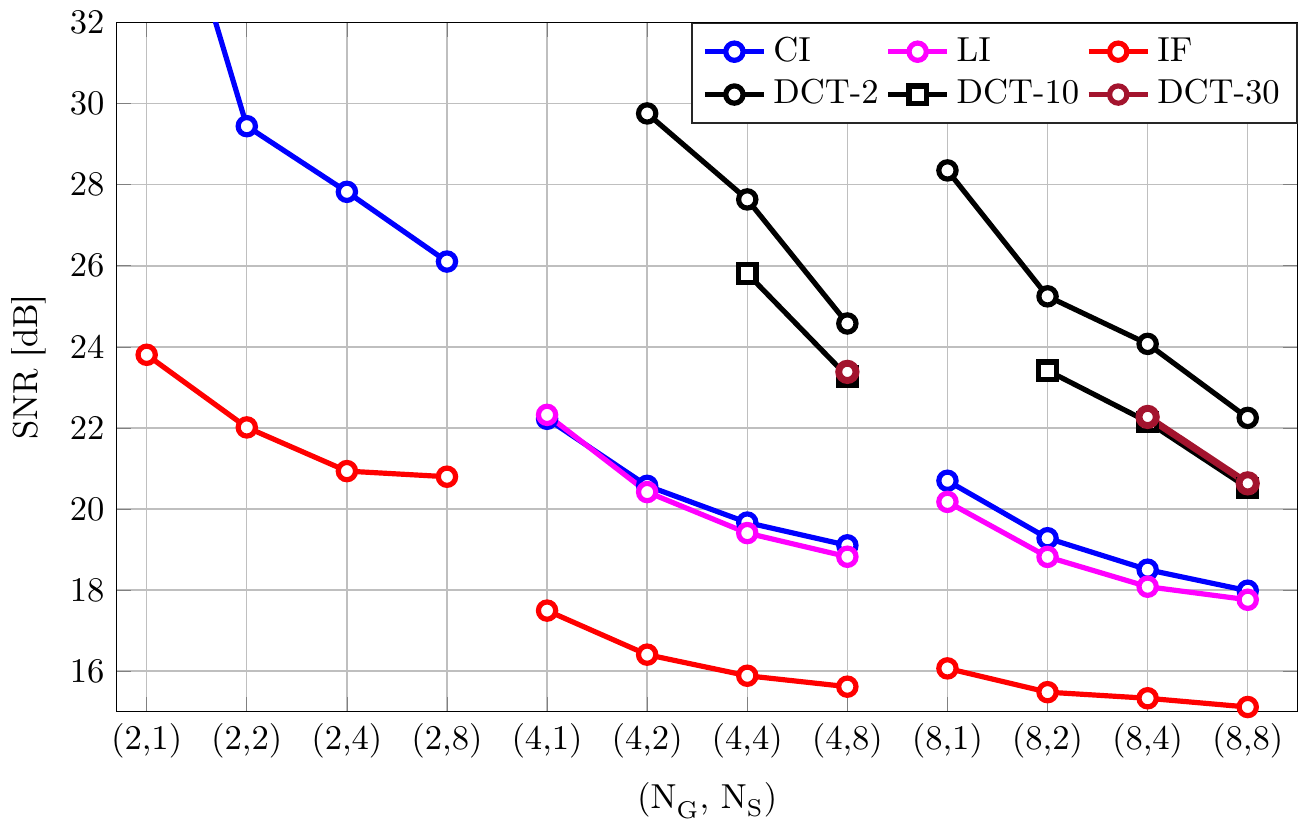}
    \caption{SNR achieved in function of $(N_G,N_S)$ to reach a target uncoded BER = $10^{-4}$ using a 16-QAM modulation and PN at the carrier frequency of 140 GHz. \vspace{0cm}}
    \label{fig:Perf1}
\end{figure}

The Fig. \ref{fig:Perf1} depicts the performance of a DFT-s-OFDM system for different ($N_G,N_S$) configurations for a target BER=$10^{-3}$.
For $N_G=2$, we observe that the CI algorithm provides better performance than LI and DCT but IF provides the best performance both in high and low density pilots. When ($N_G,N_S$) increases w.r.t the increase of inserted PTRS, the performance gap between IF and other algorithms decreases. System performance achieved by CPEE are not presented because they do not reach the target BER. It is because the average over PTRS symbols is not representative for a large DFT spread block size.

For the DCT algorithm, we observe that the system performance achieve the target at high PTRS density for DCT-($2$,$10$,$30$) but not for DCT-$1$. In the case of low PTRS density, the DCT does not achieve the target BER. Indeed, the DCT may cause poor results in two cases: i) when $N_D>K$ and in that case the condition in (\ref{eq_DCT4}) is not respected resulting in a wrong estimation of the PN; or ii) when $N_D \leq K$ but the chosen $N_D$ is not optimal or cannot properly estimate the PN. As highlighted, IF algorithm provides better performance than other algorithms for the smallest ($N_G,N_S$) configuration.

%%%%%%%%%%%%%%%%%%%%%%%%%%%%%%%%%%%%%%%%%%%%%%%%%%%%%%%%%%%%%%%%%%%%%%%%%%%%%%%%%
\subsection{Transport Block Error Rate (TBLER) Performance}
%%%%%%%%%%%%%%%%%%%%%%%%%%%%%%%%%%%%%%%%%%%%%%%%%%%%%%%%%%%%%%%%%%%%%%%%%%%%%%%%%
We evaluate the demodulation performance of the coded DFT-s-OFDM system using algorithms presented in \ref{Algo}. We choose a low density parity check (LDPC) code respecting the 5G-NR specifications \cite{3GPP3}. The PTRS distribution in Fig. \ref{fig:PTRS}-(a) is considered for coded simulations. The signal-to-noise ratio (SNR) is evaluated on the full signal bandwidth. The discrete-time PN is generated by considering the method presented in \cite{3GPP2}.

\begin{figure}[tb]
\centering
    \includegraphics[width=1.03\columnwidth]{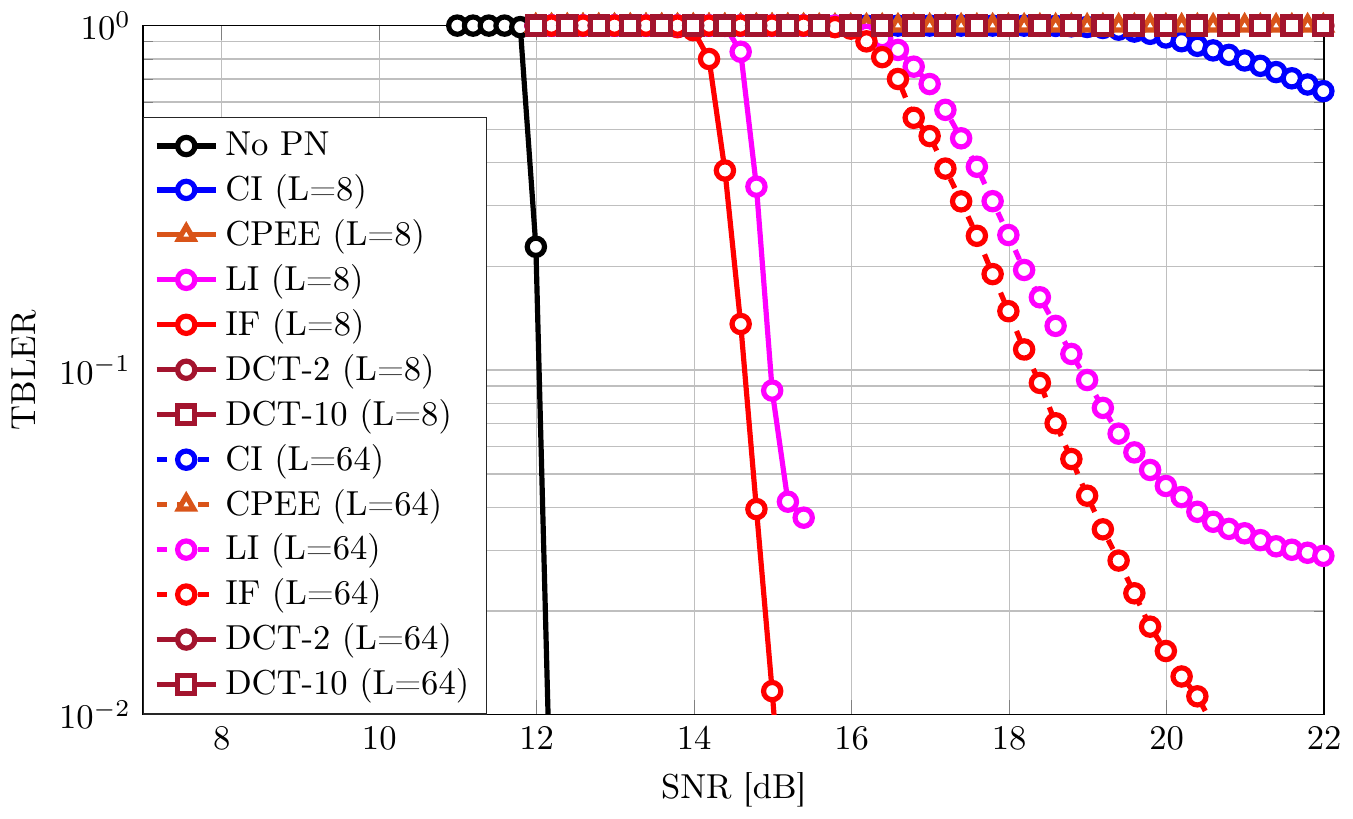}
    \caption{TBLER in function of SNR considering a $64$-QAM and $300$GHz of carrier frequency. Normal line corresponds to high PTRS density ($L$=$8$) and the dotted line to low PTRS density ($L$=$64$). \vspace{0cm}}
    \label{fig:Perf2}
\end{figure} 

Fig. \ref{fig:Perf2} presents the DFT-s-OFDM system performance considering a $64$-QAM modulation and a discrete-time PN at the carrier frequency of $300$GHz. Unlike the uncoded case, the DCT does not achieve a TBLER =$10^{-1}$. This is possibly due to the use of high order modulation order and carrier frequency. The CPEE does not provide a good estimation of PN. We observe that the IF algorithm outperforms LI, CI, CPEE and DCT algorithms in both high PTRS ($L$=$8$ \ie~$K=128$) and low PTRS ($L$=$64$ \ie~$K=16$) density. For a TBLER=$10^{-1}$, we remark respectively a SNR gain of $0.2$ dB for $L$=$8$ and $0.8$ dB for $L$=$64$ using IF compared to LI. We also observe an error floor with LI algorithm which never achieves a TBLER=$10^{-2}$. 

%%%%%%%%%%%%%%%%%%%%%%%%%%%%%%%%%%%%%%%%%%%%%%%%%%%%%%%%%%%%%%%%%%%%%%%%%%%%%%%%%%
\section{Conclusion}\label{Conclu}
%%%%%%%%%%%%%%%%%%%%%%%%%%%%%%%%%%%%%%%%%%%%%%%%%%%%%%%%%%%%%%%%%%%%%%%%%%%%%%%%%%
In this paper, we proposed a PN estimation algorithm called IF algorithm. This algorithm is based on the use of stochastic properties of PN-induced effects. We compared the proposed algorithm with existing algorithms in the state-of-the-art. We showed that the proposed algorithm outperforms all the other presented algorithms. This algorithm takes advantage of the statistical knowledge and allows good results even for low PTRS density scenario. We also showed that the IF algorithm can be implemented with 3GPP PTRS distribution.

%%%%%%%%%%%%%%%%%%%%%%%%%%%%%%%%%%%%%%%%%%%%%%%%%%%%%%%%%%%%%%%%%%%%%%%%%%%%%%%%%%
\bibliographystyle{IEEEtran}
\bibliography{IEEEabrv,reference}
%\bibliography{reference}
\end{document}